\documentclass[sigconf]{acmart}
\AtBeginDocument{%
  \providecommand\BibTeX{{%
    \normalfont B\kern-0.5em{\scshape i\kern-0.25em b}\kern-0.8em\TeX}}}

\setcopyright{acmcopyright}
\copyrightyear{2018}
\acmYear{2018}
\acmDOI{10.1145/1122445.1122456}

\acmConference[CACM'22]{Woodstock '18: ACM Symposium on Neural
  Gaze Detection}{2022}{USA}
\acmBooktitle{Woodstock '18: ACM Symposium on Neural Gaze Detection,
  June 03--05, 2018, Woodstock, NY}
\acmPrice{15.00}
\acmISBN{978-1-4503-XXXX-X/18/06}

\setcopyright{none}
\renewcommand\footnotetextcopyrightpermission[1]{}
\usepackage{subcaption}
\usepackage{makecell}
\usepackage{tabularx}
\usepackage{xcolor}
\usepackage{graphics}

\newcommand{\revision}{\color{black}}
\newcommand{\rev}[1]{\textcolor{black}{#1}}

\usepackage{todonotes}
\makeatletter
\newcommand*\iftodonotes{\if@todonotes@disabled\expandafter\@secondoftwo\else\expandafter\@firstoftwo\fi}  % defines \iftodonotes{<true>}{<false>}, thanks to https://tex.stackexchange.com/questions/126559/conditional-based-on-packageoption
\makeatother

\usepackage{listings}
\usepackage{color}
\definecolor{dkgreen}{rgb}{0,0.6,0}
\definecolor{gray}{rgb}{0.5,0.5,0.5}
\definecolor{mauve}{rgb}{0.58,0,0.82}
\definecolor{bg}{rgb}{0.9,0.9,0.9}

\usepackage{enumitem}
\setlist[itemize,1]{itemsep=1pt,partopsep=0.5pt,parsep=\parskip, topsep=2pt, leftmargin=10pt,}
\setlist[enumerate,1]{itemsep=1pt,partopsep=0.5pt,parsep=\parskip, topsep=2pt, leftmargin=10pt,}

\usepackage{layouts}
\usepackage{titlesec}
\titlespacing*{\section}{1pt}{1\baselineskip}{0.5\baselineskip}
\titlespacing*{\subsection}{1pt}{1\baselineskip}{0.5\baselineskip}
\titlespacing*{\subsubsection}{1pt}{0.8\baselineskip}{0.5\baselineskip}

\begin{document}

\title{Democratizing Domain-Specific Computing}

\author{Yuze Chi, Weikang Qiao, Atefeh Sohrabizadeh, Jie Wang, Jason Cong}
\email{{chiyuze, wkqiao2015, atefehsz, jiewang, cong}@cs.ucla.edu}
\affiliation{%
  \institution{University of California Los Angeles}
  \city{Los Angeles}
  \state{CA}
  \country{USA}
}


\begin{abstract}
    
In the past few years, domain-specific accelerators (DSAs), such as Google's Tensor Processing Units, have shown to offer significant performance and energy efficiency over general-purpose CPUs. An important question is whether typical software developers can design and implement their own customized DSAs, with affordability and efficiency, to accelerate their applications. This article presents our answer to this question.

\end{abstract}


\maketitle

\section{Introduction}\label{sec:intro}
General-purpose computers are widely used in our modern society. There were close to 24 million software programmers worldwide as of 2019 according to Statista. However, the performance improvement of general-purpose processors has slowed down significantly due to multiple reasons.
One is the end of Dennard scaling~\cite{dennard74}, which scales transistor dimensions and supply powers by 30\% every generation (roughly every two years), resulting in a $2\times$ increase in the transistor density and a 30\% reduction of the transistor delay (or improvement in the processor frequency)~\cite{borkar2011future}.
Although the transistor density continues to double per generation according to the Moore's law, the increase of the processor frequency was slowed or almost stopped with the Dennard scaling ending in the early 2000s (due to the leakage current concern).
The industry entered the era of parallelization, with tens to thousands of computing cores integrated in a single processor, and tens of thousands of computing servers connected in a warehouse-scale data center.
However, by the end of 2000s, such massively parallel general-purpose computing systems were again faced with serious challenges in terms of power supply, heat dissipation, space, and cost ~\cite{esmaeilzadeh2011dark,cong2010customizable, cacm21-thompson}.

To further advance the computing performance, \textit{customized computing} was introduced where one can adapt the processor architecture to match the computing workload for much higher computing efficiency using special-purpose accelerators~\cite{cong2010customizable, cong2012charm, brooks-islped13, dally2020domain}.
The best known customized computing example is probably the Tensor Processing Unit (TPU)~\cite{tpu} announced by Google in 2017 for accelerating machine learning workloads. Designed in 28nm CMOS technology as an application-specific integrated circuit (ASIC), TPU demonstrated $196\times$ performance/watts power efficiency advantage over the general-purpose Haswell CPU, a leading server-class CPU at that time of publication.  One significant source of energy inefficiency of a general-purpose CPU comes from its long instruction pipeline, time-multiplexed by tens or even hundreds of different types of instructions, resulting in high energy overhead (64\% for a typical superscalar out-of-order pipeline studied in ~\cite{cong2014accelerator}).
In contrast, \textit{the domain-specific accelerators (DSAs)} achieve their efficiency in the following five dimensions~\cite{dally2020domain}: (i) use of special data types and operations, (ii) massive parallelism, (iii) customized memory accesses, (iv) amortization of the instruction/control overhead, and (v) algorithm and architecture co-design. When these factors are combined, a DSA can offer significant (sometimes more than $100,000\times$) speedup over a general-purpose CPU~\cite{dally2020domain}.

Given that DSAs are domain-specific,  a key question is if a typical application developer in a given application domain can easily implement their own DSAs.  For ASIC-based DSAs, such as TPUs, there are two significant barriers.  One is the design cost.  According to McKinsey, the cost of designing an ASIC with a leading edge technology (7nm CMOS) is close to \$300M~\cite{mckinsey}, 
which is prohibitively high for most companies and developers.  The second barrier is the turnaround time.  It usually takes more than 18 months from the initial design to the first silicon, and even longer to production.  During this time, new computation models and algorithms may emerge, especially in some fast-moving application fields, making the initial design out-dated.

In light of these concerns, we think that the field-programmable gate-arrays (FPGAs) offer an attractive alternative for DSA implementation.  Given its programmable logics,  programmable interconnects, and customizable building blocks (BRAMs and DSPs), an FPGA can be customized to implement a DSA without going through a lengthy fabrication process and can be reconfigured to a new design in a matter of seconds. 
Moreover, FPGAs have become available in the public clouds, such as Amazon AWS F1~\cite{amazon-f1} and Nimbix~\cite{nimbix}. One can create their own DSAs on the FPGA and use it at a rate of \$1-2/hour to accelerate the desired applications, even if FPGAs are not available in the local computing facility.   Because of its affordability and fast turnaround time, we think that FPGAs offer the promise of \textit{democratization of customized computing}, allowing  millions of software developers to create their own DSAs on FPGAs for  performance and energy efficiency.  
\rev{Although a DSA implemented on an FPGA is less efficient than the one on an ASIC due to the lower circuit density and clock frequency, it can still deliver tens or hundreds of times better efficiency compared to CPUs (as shown in Section~\ref{sec:motivation}).}

However, to achieve the true democratization of customized computing, a convenient and efficient compilation flow needs to be provided for a typical performance-oriented software programmer to create a DSA on an FPGA, either on premise or in the cloud.  Unfortunately, this has not been the case. FPGAs used to be designed with hardware description languages, such as Verilog and VHDL, known only to the circuit designers.  In the past decade, FPGA vendors introduced the high-level synthesis (HLS) tools to compile C/C++/OpenCL programs to FPGAs.  Although these HLS tools raise the level of design abstraction, they still require a significant amount of hardware design knowledge, expressed in terms of pragmas, to define how computation is parallalized and/or pipelined, how data are buffered, how memory is partitioned, etc. As shown in Section~\ref{sec:hls-limitation}, the performance of a DSA design can vary from being $108\times$ slower (without performance-optimizing pragmas) than a CPU to $89\times$ faster with proper optimization. 
But such architecture-specific optimization is often beyond the reach of an average software programmer.

In this paper, we highlight our research on democratizing customized computing by providing highly effective compilation tools for creating customized DSAs on FPGAs. It builds on top of the HLS technology, but 
\rev{greatly reduces (or completely eliminates in some parts/sections) the need for} pragmas for hardware-specific optimization. This is achieved by high-level architecture-guided optimization and automated design space exploration. 
Section~\ref{sec:motivation} uses two examples to showcase the efficiency of DSAs on FPGAs over the general-purpose CPUs. Section~\ref{sec:hls-limitation} discusses challenges of programming an FPGA and Section~\ref{sec:automation} reviews our proposed solutions using architecture-guided optimization, such as systolic array \rev{(Section~\ref{sec:autosa})} or stencil computation \rev{(Section~\ref{sec:soda})}, automated design space exploration \rev{for general applications (Section~\ref{sec:autodse}), and raising the abstraction level to domain-specific languages (DSLs) (Section~\ref{sec:heterocl})}. Section~\ref{sec:conclusion} concludes the paper and discusses future research directions. The focus of this paper is on creating new DSAs (on FPGAs) instead of programming existing DSAs, such as GPUs and TPUs, which are also highly efficient for their target workloads. Some of the techniques covered in this paper can be extended for the latter, such as supporting systolic arrays or stencil computation on GPUs ~\cite{wang2017communication, sc19-stencil}. This paper is based on the keynote speech given by one of the co-authors at the 2021 International Parallel and Distributed Processing Symposium (IPDPS)~\cite{cong-ipdps}.

\section{Promise of Customizable Domain-Specific Acceleration} \label{sec:motivation}

In this section, we highlight two DSAs on FPGAs targeting sorting and deep learning applications to demonstrate the power of customizable domain-specific acceleration.

\subsection{High-Performance Sorting}
Our first example to showcase an FPGA-based DSA is accelerated sorting, which is a fundamental task in many big data applications. One of the most popular sorting algorithms for large-scale sorting is recursive merge sort, given its optimal computation and I/O communication complexity. However, \rev{the slow sequential merging steps usually limit the performance of this recursive approach. Although parallel merging is possible, this comparison-based approach often comes with high overhead on CPUs and GPUs and limits the throughputs, especially in the last merging stages.}

\begin{figure}[!h]
    \centering
    \includegraphics[width=0.95\columnwidth]{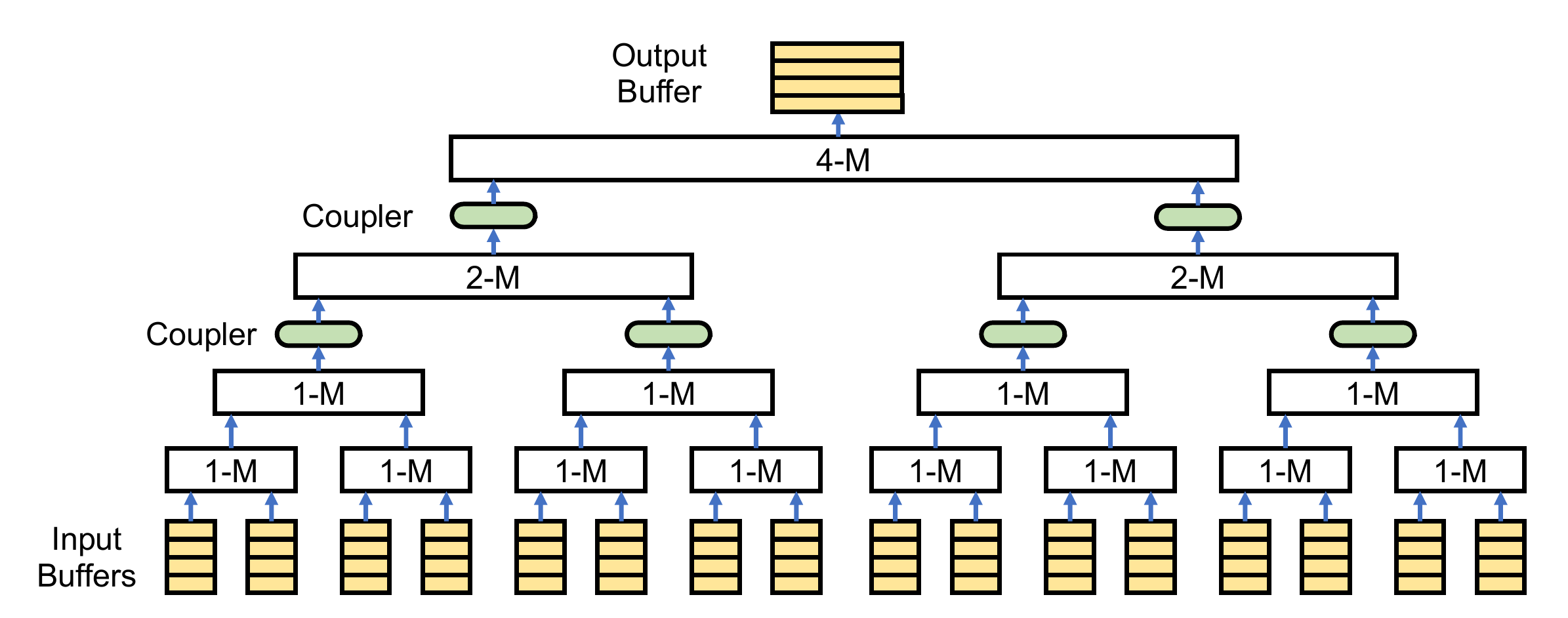}
    \caption{A merge tree that merges 16 input streams simultaneously and outputs 4 elements per cycle.}
    \label{fig:merge_tree}
\end{figure}

FPGAs have abundant on-chip computing resources (e.g., LUTs and DSPs) and memory resources (e.g., registers and BRAM slices) available. One can achieve impressive performance speedup by implementing the recursive merging flow into a tree-based customizable spatial architecture~\cite{isca20,fccm21} as in Figure~\ref{fig:merge_tree}. A merge tree is  uniquely defined by the number of leaves and the throughput at the root. The basic building block in the hardware merge tree is  a $k$-Merger (denoted as $k$-M in Figure~\ref{fig:merge_tree}), which is a customized logic that can merge two sorted input streams at a rate of $k$ elements per cycle in a pipelined fashion. Using a combination of such mergers with different \rev{throughputs}, one can build a customized merge tree with an optimized number of leaves and root throughput for a given sorting problem and memory configuration. 

\begin{figure}[!h]
    \centering
    \includegraphics[width=0.95\columnwidth]{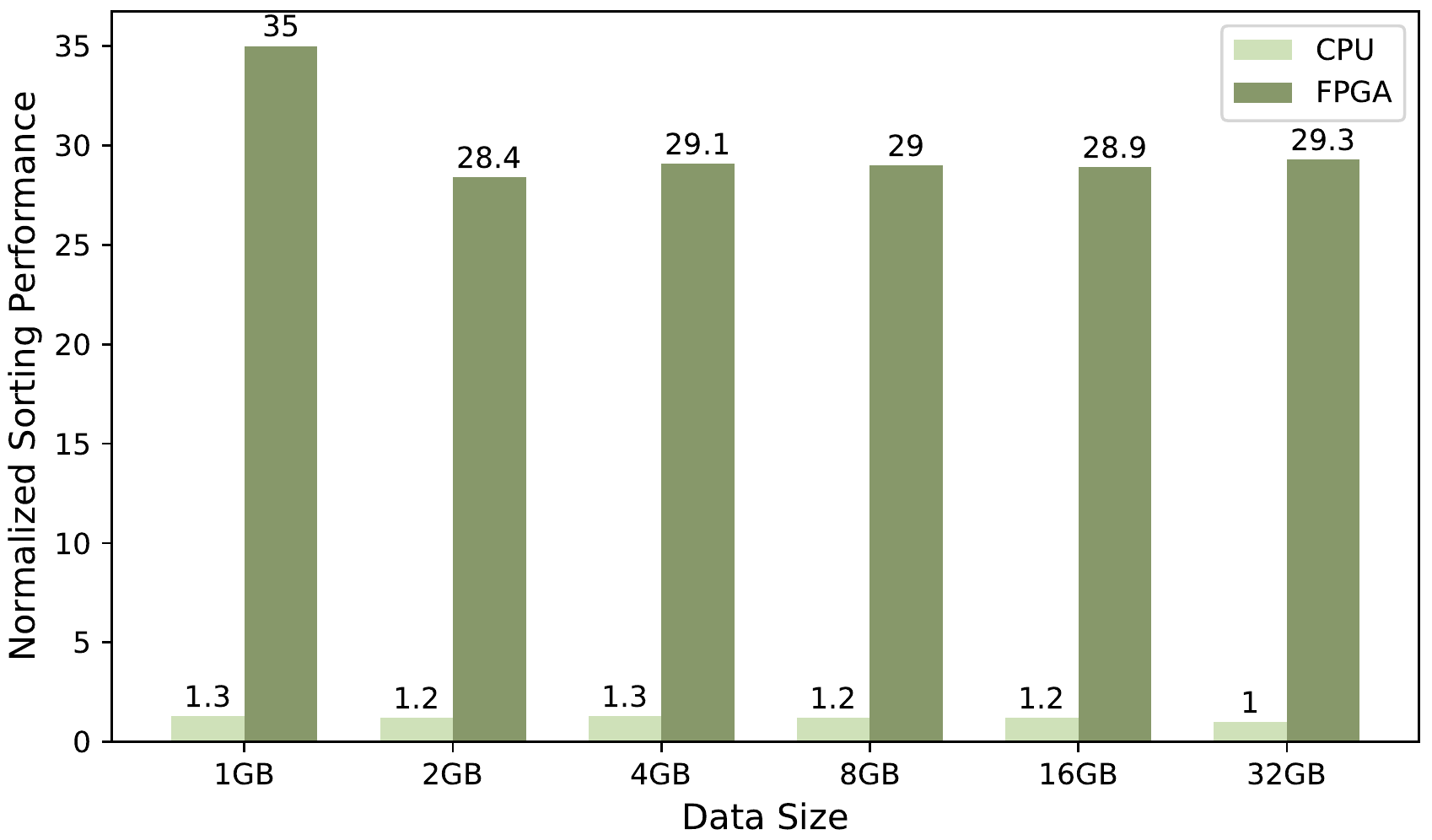}
    \caption{Comparison of the sorting performance using CPUs and FPGA accelerators.}
    \label{fig:sorting_result}
\end{figure}

Figure~\ref{fig:sorting_result} shows the speedup that the customized merge tree accelerator on the AWS F1 FPGA instance achieves over a 32-thread Intel Xeon CPU implementation (the absolute performance of baseline is 0.21 GB/s). Part of this large efficiency gain arises from the common advantages of DSAs as discussed in Section~\ref{sec:intro}: e.g., (i) specialized data type and operations: the hardware mergers support any key and value width up to 512 bits; (ii) massive data parallelism: the customized merge tree is able to merge 64 input streams concurrently and output 32 integer elements every cycle; (iii) optimized on-chip memory: we optimize each input buffer preparing the input stream to its corresponding tree leaf to have enough space to hide the DRAM access latency. However, there are additional FPGA-related features that enable us to achieve the high efficiency.
\begin{itemize}
    \item \textbf{Design choices tailored to hardware constraints:}  We can tailor the customizable merge-tree-based architecture according to the given workload and operating environment. For example, since the available DRAM bandwidth on AWS F1 is 32 GB/s for concurrent read and write, we tune the tree root throughput to be the same amount and select the maximum number of tree leaves that fit into the on-chip resources. This minimizes the number of passes needed for merging.
    \item \textbf{Flexible reconfigurability:} FPGAs have a unique feature to support hardware reconfigurability at the runtime. When sorting multiple large datasets from different domains, one may pre-generate multiple merge-tree configurations, each optimized for its own use case (e.g., data size, key/data width). Then, we reprogram the FPGA at the run time to adapt our accelerator to the changing sorting demands. Also, when sorting terabyte-scale data stored on SSDs, the merge sort needs to go through two phases: 1) to merge the data up to the DRAM size, 2) to finally merge the data onto the SSD. We show that designers can reconfigure the FPGA to switch between these two phases efficiently in~\cite{fccm21}.
\end{itemize}

\subsection{DNN Accelerators}
Our second example is FPGA-based acceleration of deep neural networks (DNNs). DNNs have been widely used for many artificial intelligence (AI) applications ranging from computer vision and speech recognition to robotics, due to their greatly improved accuracy and efficiency. 
One of the earliest, also probably the most cited FPGA-based deep learning accelerator is the one published in early 2015~\cite{fpga15-dnn}. It was developed using HLS to accelerate multi-layer convolution neural networks (CNN). The whole system was implemented on a single Xilinx Virtex-7 485T FPGA chip with a DDR3 DRAM. It demonstrated close to $5\times$ speedup and $25\times$ energy reduction compared to a 16-thread CPU implementation. Utilizing HLS, it was able to explore over 1,000 accelerator design choices based on the roofline model and converged to a solution that is optimized both for computation and communication. Several graduate students carried out this implementation in less than six months and completed it almost two years ahead of Google's announcement of TPU~\cite{tpu}, which was done by a much larger design team.

Microsoft also designed an FPGA-based DNN accelerator, named Brainwave Neural Processing Unit (NPU)~\cite{brainwave}, at a much larger scale and has widely deployed Brainwave NPUs in its cloud production. \rev{Table~\ref{tbl:BW_compare} summarizes the hardware specifications and benchmark results of Brainwave NPU re-implemented on Intel S10 NX and NVIDIA GPUs~\cite{fpt20}. The results show that FPGAs can not only achieve an order of magnitude better performance than GPUs for low-batch inferences, but also compete in high-batch inference cases. On the other hand, although there is a performance gap between FPGA-based NPUs and hardened NPUs~\cite{asic20}, the re-configurable nature of FPGAs allows designers to quickly adapt their designs to the emerging deep learning algorithms. The advantages of FPGA-based DNN accelerators are listed below:}

\begin{itemize}
    \item \textbf{Customized data type:} Deep neural networks are highly compressible in data types since using a low-precision customized floating point format has negligible impact on accuracy. For example, the Brainwave NPU employs a narrow floating point which contains 1-bit sign, 5-bit exponent and 2-bit mantissas. \rev{Although some customized bit-width (e.g., 16-bit and 8-bit) support is added in the latest GPUs, the FPGAs demonstrate the flexibility to go down to ultra-low bit width (1-2 bits) with dynamic customization \cite{fracbnn}, as such narrow-precision data types can be mapped efficiently onto LUTs on FPGAs.} 
    \item \textbf{Synthesizable parameters:} The Brainwave NPU allows for several parameters to be changed such as data type, vector size, number of data lanes, and size of the matrix-vector tile engine during the synthesis. As a result, designers can specialize its architecture for various DNN models without expensive hardware updates. \rev{This gives FPGAs a distinct advantage over ASICs in terms of the costs and development cycles, as today's deep learning models are still evolving.}%This is extremely important for real-world cloud service providers as the models are still evolving and redesigning an ASIC may not be affordable.
    \item \textbf{\rev{Low system overhead latency:}} Part of the FPGA fabric supports efficient packet processing schemes, allowing data to be offloaded onto the accelerator with extremely low latency: e.g., the FPGAs are programmed to process network packets from remote servers with little overhead and feed the data into the Brainwave NPU on the same die at the line rate. Such low-latency data offloading ensures that the users can get the acceleration result in real time, whether the data are from the edge or the cloud. It also allows a large DNN model to be decomposed and implemented on multiple FPGAs with very low latency. 
\end{itemize}

\begin{table}[!htb]
% \footnotesize
\centering
\caption{\rev{Hardware specifications and inference performance comparison (GRUs and LSTMs) on FPGAs and GPUs~\cite{fpt20}.}}
\label{tbl:BW_compare}
\resizebox{\columnwidth}{!}{%
\begin{tabular}{cccc}
\hline
\multicolumn{1}{|c|}{Hardware} & \multicolumn{1}{c|}{Nvidia T4} & \multicolumn{1}{c|}{Nvidia V100} & \multicolumn{1}{c|}{Intel S10 NX} \\ \hline\hline
\multicolumn{1}{|c|}{Peak FP16 TOPS} & \multicolumn{1}{c|}{65} & \multicolumn{1}{c|}{125} & \multicolumn{1}{c|}{143} \\ \hline
\multicolumn{1}{|c|}{Peak INT8 TOPS} & \multicolumn{1}{c|}{130} & \multicolumn{1}{c|}{63} & \multicolumn{1}{c|}{143} \\ \hline
\multicolumn{1}{|c|}{On-Chip Mem. (MB)} & \multicolumn{1}{c|}{10} & \multicolumn{1}{c|}{16} & \multicolumn{1}{c|}{16} \\ \hline
\multicolumn{1}{|c|}{Process} & \multicolumn{1}{c|}{TSMC 12nm} & \multicolumn{1}{c|}{TSMC 12nm} & \multicolumn{1}{c|}{Intel 14nm} \\ \hline\hline
\multicolumn{1}{|c|}{Speedup at batch-3} & \multicolumn{1}{c|}{1$\times$} & \multicolumn{1}{c|}{2.4$\times$} & \multicolumn{1}{c|}{22.3$\times$} \\ \hline
\multicolumn{1}{|c|}{Speedup at batch-6} & \multicolumn{1}{c|}{1$\times$} & \multicolumn{1}{c|}{2.1$\times$} & \multicolumn{1}{c|}{24.2$\times$} \\ \hline
\multicolumn{1}{|c|}{Speedup at batch-32} & \multicolumn{1}{c|}{1$\times$} & \multicolumn{1}{c|}{2.5$\times$} & \multicolumn{1}{c|}{5.0$\times$} \\ \hline
\multicolumn{1}{|c|}{Speedup at batch-256} & \multicolumn{1}{c|}{1$\times$} & \multicolumn{1}{c|}{2.3$\times$} & \multicolumn{1}{c|}{1.6$\times$} \\ \hline
\end{tabular}%
}
\end{table}

\rev{The recent effort by Amazon on advanced query acceleration of its Redshift database~\cite{aqua} is another good example of FPGA acceleration in datacenters.  However, wider adoption of FPGA acceleration has been constrained by the difficulty of FPGA programming, which is the focus of this paper.}

\section{FPGA Programming Challenges} \label{sec:hls-limitation}
So far it has not been easy for a typical performance-oriented CPU programmer to create their own DSAs on an FPGA to achieve the performance gain demonstrated in the preceding section, despite the recent progress on high-level synthesis (HLS).

\begin{lstlisting}[language=C,caption=HLS C code snippet of CNN. The \texttt{scop} pragma will be used for systolic compilation in Section~\ref{sec:autosa} \label{code:cnn_hls},
    float,floatplacement=H]
void CNN(float In[B][I][H+P-1][W+Q-1], 
          float W[O][I][P][Q], 
          float Out[B][O][H][W]) {
  // Use the pragma below to annotate the start of the code region to be mapped to a systolic array in AutoSA
  #pragma scop 
  for (int b = 0; b < B; b++)
    for (int o = 0; o < O; o++)
      for (int h = 0; h < H; h++)
        for (int w = 0; w < O; w++) {
          Out[b][o][h][w] = 0;
          for (int i = 0; i < I; i++)
            for (int p = 0; p < P; p++)
              for (int q = 0; q < O; q++)
                Out[b][o][h][w] += W[o][i][p][q] * In[b][i][h+p][w+q];
       }
    // Use the pragma below to annotate the end of the code region to be mapped to a systolic array in AutoSA   
    #pragma endscop
\end{lstlisting}
HLS allows a designer to start with C/C++ behavior description instead of the low-level cycle-accurate register-transfer level (RTL) description to carry out FPGA designs, which significantly shortens the turnaround times and reduces the FPGA development cycle ~\cite{cong11, zhang2008autopilot, cong2022fpga}. 
\rev{As a result,  most FPGA vendors have commercial HLS tools, e.g., Xilinx Vitis~\cite{vitis-platform} and Intel FPGA SDK for OpenCL~\cite{intel-sdk}.} 
However, even though HLS is increasingly employed by hardware designers, software programmers still find it challenging to use the existing HLS tools. 
For example, Code~\ref{code:cnn_hls} shows the C code for one layer of CNN. When synthesized with the Xilinx HLS tool, the resulting microarchitecture is, in fact, $108\times$ slower than a single-core CPU. As explained in~\cite{CPP}, this is because the derived microarchitecture has the following inefficiencies which limit its performance:
\begin{itemize}
    \item \textbf{Inefficient off-chip communication:} Although the bandwidth of the off-chip memory can support fetching 512-bit data at a time, the HLS solution uses only 32 bits of the available bus. This is because the input arguments of the function in Code~\ref{code:cnn_hls} (lines 1-3), which create interfaces to the off-chip memory, use 32-bit floating-point data type.
    \item \textbf{No data caching:} Lines 10 and 14 of Code~\ref{code:cnn_hls} access the off-chip memory directly. Although this type of memory access will be cached automatically on a CPU, they will not be cached on an FPGA by default. Instead, the designer must explicitly specify which data need to be reused using the on-chip memories (BRAM, URAM, or LUT).
    \item \textbf{No overlap between communication and computation:} A load-compute-store pipelined architecture is necessary to achieve good computation efficiency as done in most CPUs. However, this is not created automatically by the HLS tool based on the input C code.
    \item \textbf{Sequential loop scheduling:} The HLS tools require the designer to use synthesis directives in the form of pragmas to specify where to apply parallelization and/or pipelining. In their absence, everything will be scheduled sequentially.
    \item \textbf{Limited number of ports for on-chip buffers:} The default on-chip memory (i.e., BRAM) has one or two ports.  Without proper array partitioning, it can greatly limit the performance since it restricts the amount of parallel accesses to the on-chip buffers, such as the \textit{Out} buffer in Code~\ref{code:cnn_hls}.
\end{itemize}

Fortunately, these shortcomings are not fundamental limitations. They only exist because the HLS tools are designed to generate the architecture based on specific C/C++ code patterns. As such, we can resolve all of them and get to a 9,676$\times$ speedup. %which results in the micro-architecture in Figure~\ref{fig:cnn-arch-optimized}.
To achieve this, we first saturate the off-chip memory's bandwidth by packing 16 elements and creating 512-bit data for each of the interface arguments. We then explicitly define our caching mechanism and create load-compute-store stages to decouple the computation engine and the data transfer steps, so that the compute engine works only with on-chip memory. Finally, we exploit \texttt{UNROLL} and \texttt{PIPELINE} pragmas to define the parallelization opportunities. We also use \texttt{ARRAY\_PARTITION} pragmas as needed, which brings the total number of pragmas to 28 and the lines of codes to 150, to enable parallel accesses to the on-chip memory by creating more memory banks (ports). %In total, the optimized code utilizes 28 pragmas.
Table~\ref{tbl:comparison-cnn} compares the two microarchitectures in terms of their number of resources, global memory bandwidth, and performance when we map them to Xilinx Virtex Ultrascale+ VCU1525.
\begin{table}[!htb]
    \centering
    \resizebox{\columnwidth}{!}{
    \begin{tabular}{|c|c|c|c|c|c|c|c|}
         \hline
         %Architecture & \makecell{LUT / FF / BRAM / DSP \\ (\%)} & \makecell{Used DRAM \\ BW (Bits)} & Speedup \\
         %\hline \hline
         Architecture & BRAM & DSP & \makecell{Used DRAM BW (Bits)} & Speedup \\ \hline \hline
         %\hline \hline
         %Baseline HLS & 1.06 /	0.37 /	6.90 /	0.10 & 32 & $1\times$ \\
         Baseline HLS & 2.1\% &	0.1\% & 32 & $1\times$ \\
         \hline
         %Manually Optimized HLS & 54.24 / 33.48 / 78.01 / 47.21 & 512 & $7041\times$ \\
         Manually Optimized HLS & 78.0\% & 47.2\% & 512 & $9,676\times$ \\
         \hline
    \end{tabular}
    }
    \caption{Microarchitecture comparison for the naive CNN code (Code~\ref{code:cnn_hls}) and its optimized version.}
    \label{tbl:comparison-cnn}
\end{table}

\section{DSA Design Automation Beyond HLS} \label{sec:automation}
\begin{figure}[!h]
    \centering
    \includegraphics[width=0.95\columnwidth]{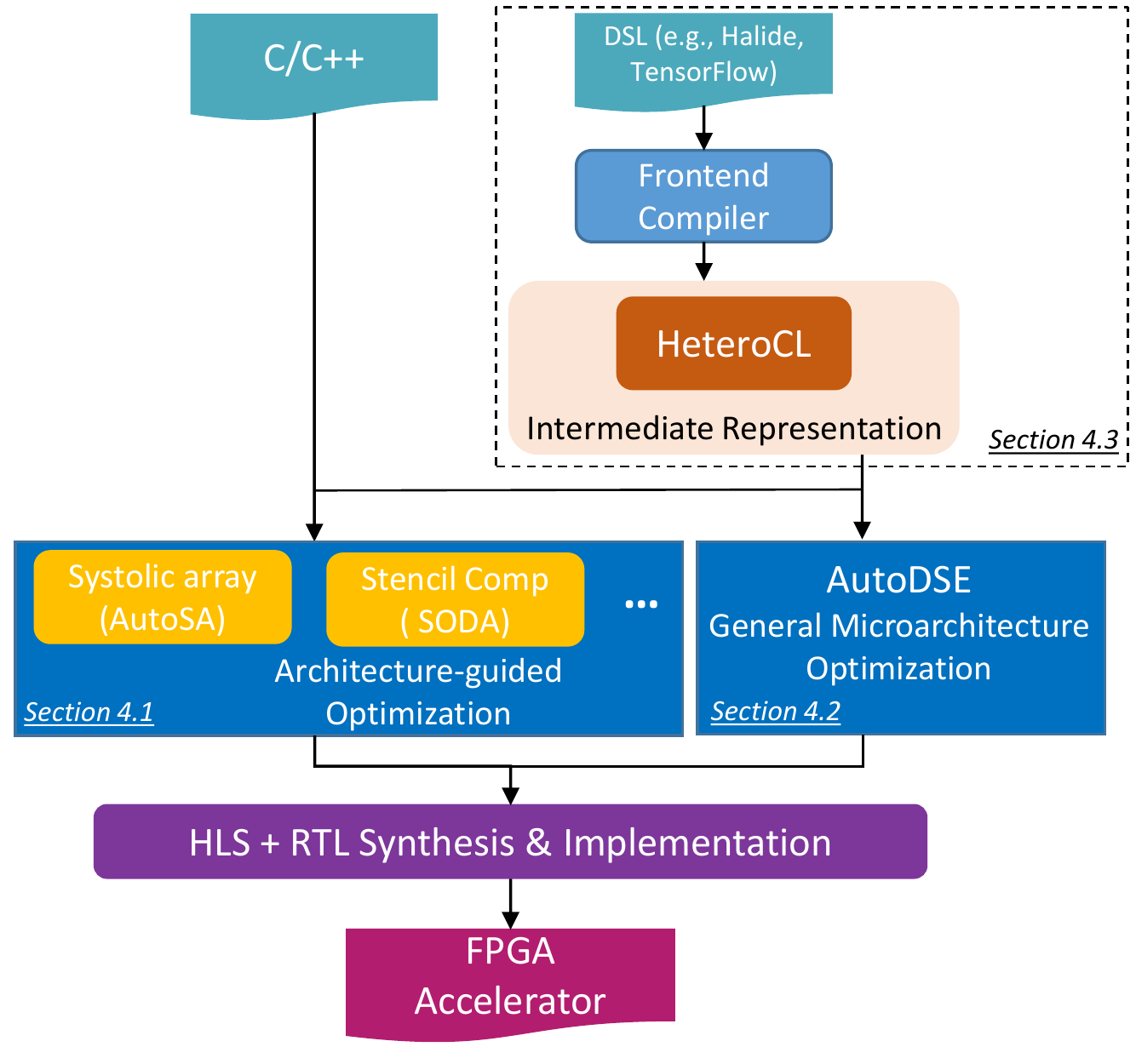}
    \caption{Overview of our approaches.}
    \label{fig:overview}
\end{figure}

To overcome the FPGA programming challenges discussed in the preceding section, in this section, we highlight our software-programmer-friendly compilation solutions for FPGAs. \rev{Figure~\ref{fig:overview} details our overall compilation flow from C, C++, or DSLs to FPGA acceleration.}
Our solutions include: using architecture-guided optimization (Section~\ref{sec:architecture-guided-optimation}), such as systolic array or sliding window-based architecture for stencil applications, automated design space exploration (Section~\ref{sec:autodse}), and domain-specific language (Section~\ref{sec:heterocl}).

\subsection{Architecture-Guided Optimization}

\label{sec:architecture-guided-optimation}
One of the challenges of existing HLS tools is that many pragmas are needed to specify a complex microarchitecture, such as a systolic array,  for efficient implementation. \rev{Instead, we allow the programmer to simply mark the section of code suitable for a certain microarchitecture pattern 
and let the tool to automatically generate complex and highly optimized HLS code for the intended microarchitecture. This is called \textit{architecture-guided optimization}}.  In this section, we showcase two examples -- compilations for systolic arrays and stencil computations.

\subsubsection{Automated systolic arrays compilation:}
\label{sec:autosa}

Systolic array architecture consists of a grid of simple and regular processing elements (PE) which are linked through local interconnects. With the modular design and local interconnects, we can easily scale out this architecture to deliver a high performance while achieving a high energy efficiency at the same time. One of the representative examples of this architecture is the Google TPU~\cite{tpu}. It implements systolic arrays as the major compute unit to accelerate the matrix operations in the machine learning applications.

On the downside, designing a high-performance systolic array can be a challenging task. It requires the expert knowledge of both the target application and the hardware. Specifically, designers need to identify the systolic array execution pattern from the application, transform the algorithm to describe a systolic array, write the hardware code for the target platform, and tune the design to achieve the optimal performance. Each step will take significant efforts, raising the bar to reap the benefits of such an architecture. For example, a technical report from Intel~\cite{hongbo_arxiv} mentioned that such a development process will take months of efforts even for industry experts.

\begin{figure}[!h]
    \centering
    \includegraphics[width=0.9\columnwidth]{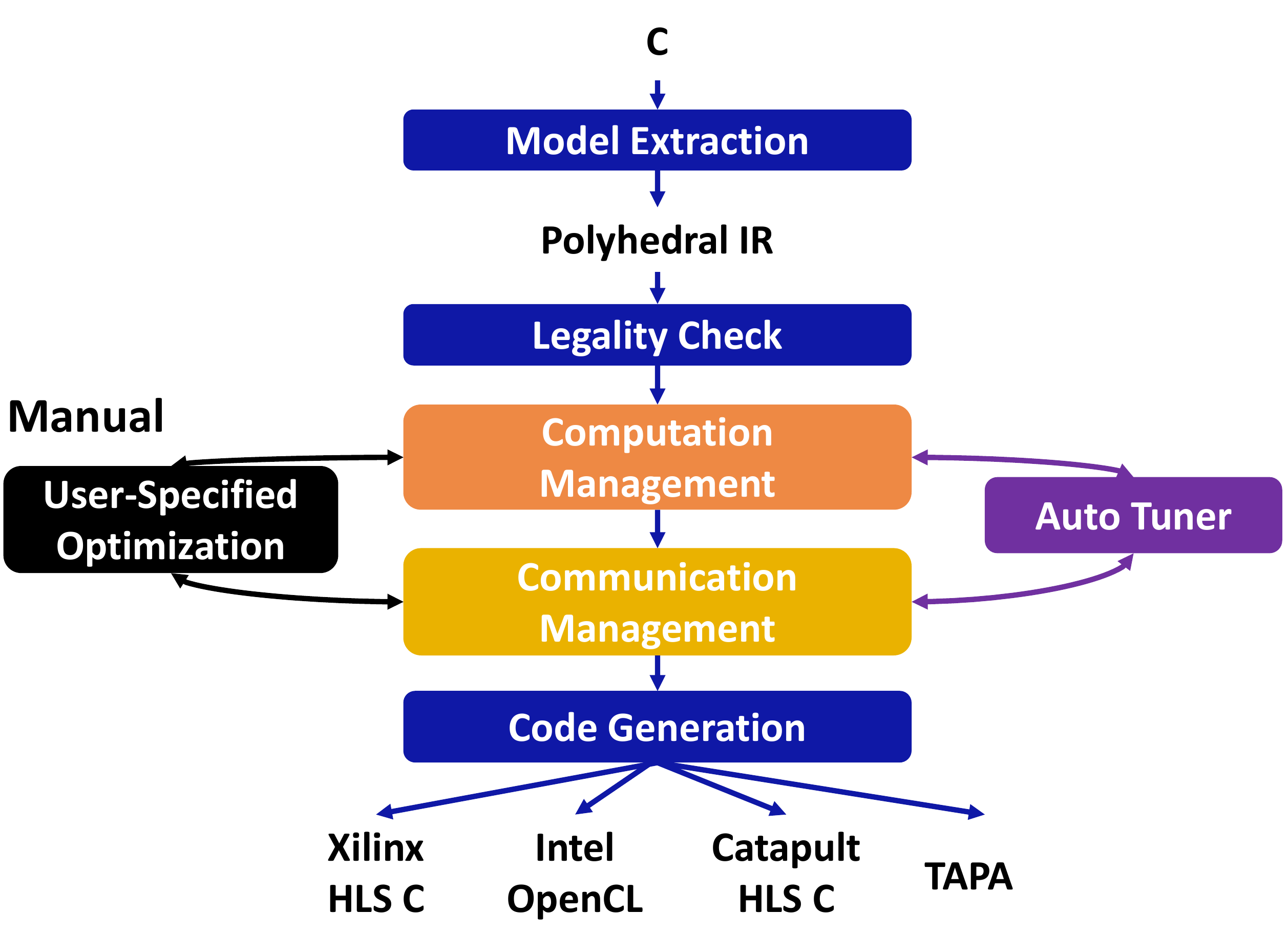}
    \caption{Compilation flow of AutoSA.}
    \label{fig:autosa_compile_flow}
\end{figure}

To cope with this challenge, we propose an end-to-end compilation framework, \textit{AutoSA}~\cite{autosa}, to generate systolic arrays on FPGA.
Figure~\ref{fig:autosa_compile_flow} depicts the compilation flow of AutoSA.
AutoSA takes a C code as the input that describes the target algorithm to be mapped to the systolic arrays. This code is then lowered to the polyhedral IR. AutoSA uses the polyhedral model~\cite{isl}, which is a mathematical compilation framework for loop nest optimization. 
\rev{AutoSA checks if the input program can be mapped to a systolic array (\textit{Legality Check})}. After that, it applies a sequence of hardware optimizations to construct and optimize the PEs (in the stage of \textit{Computation Management}) and the on-chip I/O networks for transferring the data between PEs and the external memory (\textit{Communication Management}). 
AutoSA introduces a set of tuning knobs that can either be changed manually or set by an auto-tuner.
The output of this compiler is a systolic array design described in the target hardware language. At present, we support four different hardware back-ends, including Xilinx HLS C, Intel OpenCL, Catapult HLS, and TAPA~\cite{chi2021extending}.

\begin{figure}[!h]
    \centering
    \includegraphics[width=0.5\columnwidth]{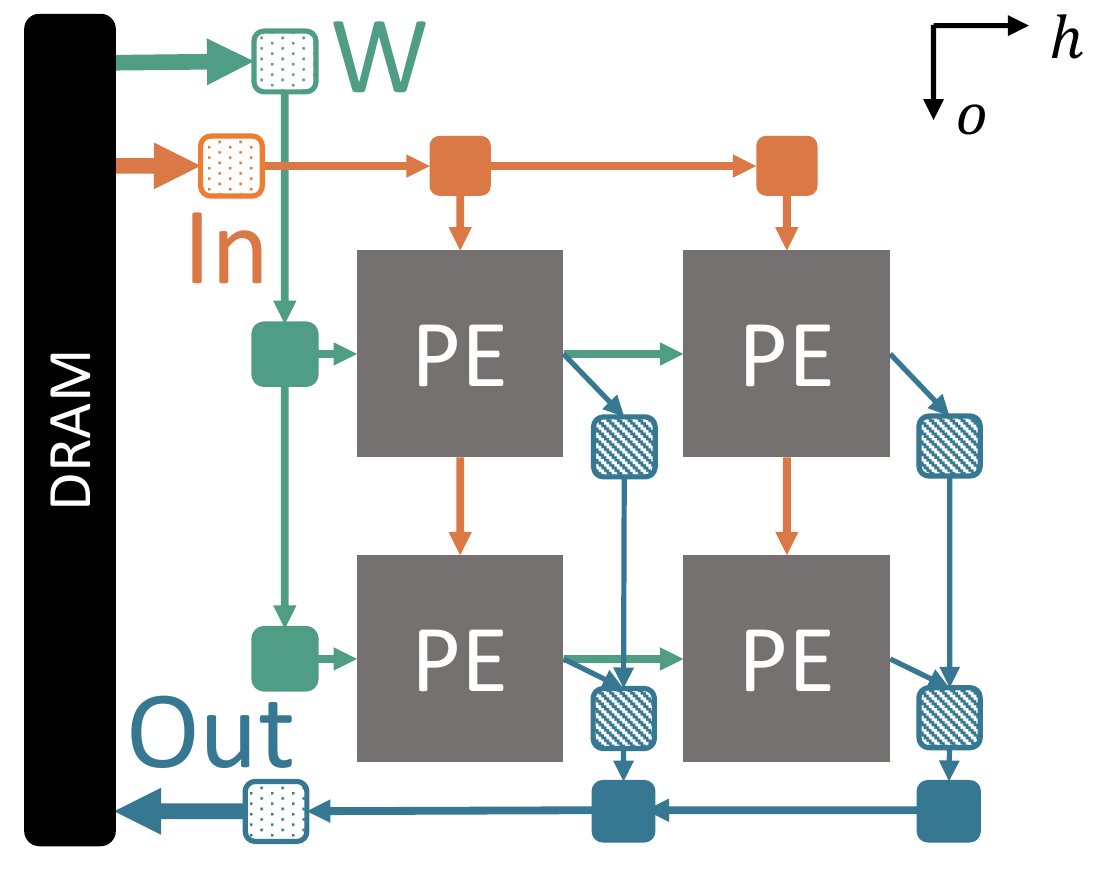}
    \caption{Architecture of a 2D systolic array for CNN.}
    \label{fig:autosa_cnn_arch}
\end{figure}

With the help of AutoSA, we could now easily create a high-performance systolic array for CNN as mentioned in the previous section. 
\rev{AutoSA requires minimal code changes to compile.
Designers only need to annotate the code region to be mapped to systolic arrays with two pragmas (Lines 5 and 17 in Code~\ref{code:cnn_hls}).}
Figure~\ref{fig:autosa_cnn_arch} shows the architecture of the systolic array with the best performance generated by AutoSA. For CNN, AutoSA generates a 2D systolic array by mapping the output channel $o$ and the image height $h$ to the spatial dimensions of the systolic array. Input feature maps $In$ are reused in-between PEs vertically, weights $W$ are reused across PEs horizontally, and output feature maps $Out$ are computed inside each PE and drained out through the on-chip I/O network. For the same CNN example as shown in Section~\ref{sec:hls-limitation}, AutoSA-generated systolic array achieved a 2.3$\times$ speedup over the HLS-optimized baseline. This is accomplished by a higher resource utilization and computation efficiency. The regular architecture and local interconnects make systolic array scalable to fully utilize the on-chip resource. Furthermore, this architecture exploits \rev{a high level} of data reuse from the application that balances the computation and communication, resulting in a high computation efficiency.  Table~\ref{tbl:comparison-cnn-autosa} compares the details of the AutoSA-generated design with the HLS-optimized baseline. \rev{Moreover, two high-level pragmas in  Code~\ref{code:cnn_hls} replaces 28 low-level pragmas in the manual HLS design. Such significant low-level HLS pragma reduction is consistently observed with the tools introduced in the later sections as well.}
%\todo{how do we name the design in the previous section? manually optimized HLS design? Add the DSP efficiency? Use a layer from existing CNNs?}

\begin{table}[!htb]
    \centering
    \resizebox{\columnwidth}{!}{
    \begin{tabular}{|c|c|c|c|c|c|c|c|}
         \hline
         %Architecture & \makecell{LUT / FF / BRAM / DSP \\ (\%)} & \makecell{Used DRAM \\ BW (Bits)} & Speedup \\
         %\hline \hline
         Architecture & BRAM & DSP & \makecell{Used DRAM BW (Bits)} & Speedup \\ \hline \hline
         %HLS-Optimized & 54.24 / 33.48 / 78.01 / 47.21 & 512 & $1.0\times$
         Manually Optimized HLS & 78.0\% & 47.2\% & 512 & $1.0\times$\\
         \hline
         AutoSA &  93.3\% & 90.1\% & 512 & 2.3$\times$ \\
         \hline
    \end{tabular}
    }
    \caption{Micro-architecture comparison for manually-optimized HLS design and AutoSA-generated design.}
    \vspace{-0.1in}
    \label{tbl:comparison-cnn-autosa}
\end{table}

\rev{Automatic systolic array synthesis has been an important research topic for decades. AutoSA represents the state-the-of-art effort along this direction. At present, AutoSA targets only FPGAs. One recent work, Gemmini~\cite{gemmini}, generates systolic arrays for deep learning applications on ASICs. Gemmini framework employs a fixed hardware template for generating systolic arrays for matrix multiplication. The general mapping methodology based on the polyhedral framework in AutoSA can be applied to Gemmini to help further improve the applicability of Gemmini framework.}

\subsubsection{Automated stencil compiler:}

\label{sec:soda}

%\hfill\\

\begin{figure}[!ht]
  \centering
  \includegraphics[trim=10mm 5mm 10mm 5mm, clip, width=\linewidth]{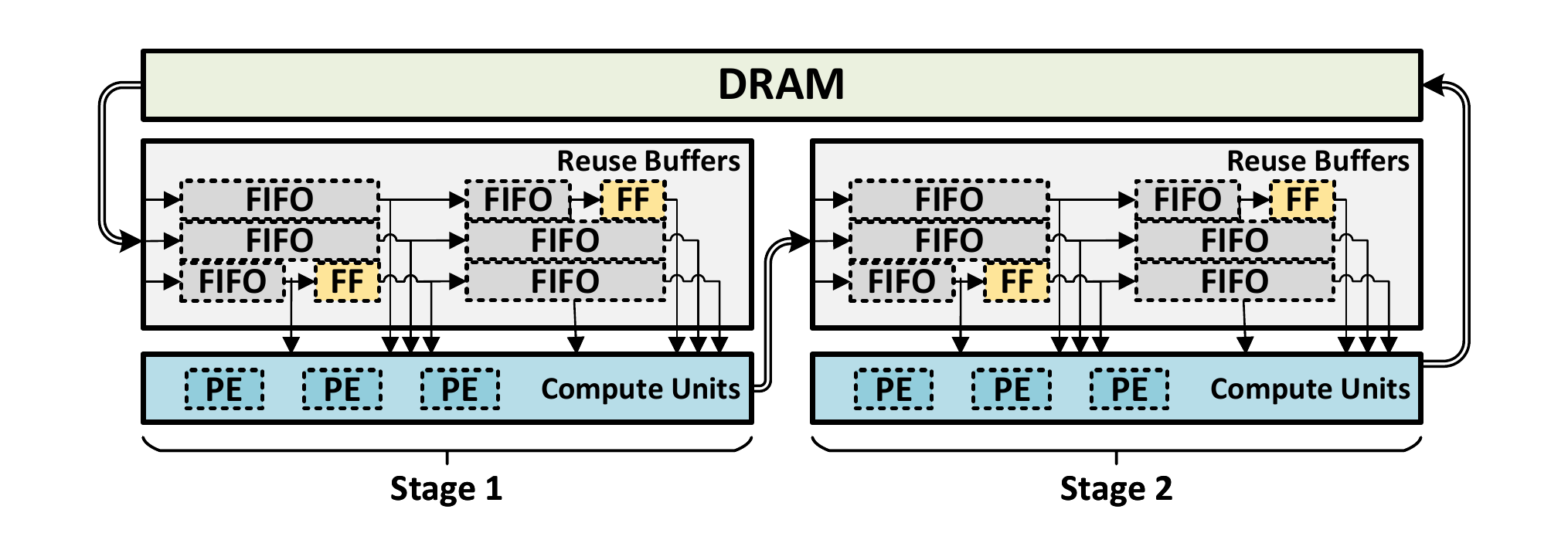}
  \caption{SODA microarchitecture overview.}
  \label{fig:soda-overview}
\end{figure}

Our second example for architecture-guided optimization is for the stencil computation, which utilizes a sliding window over the input array(s) to produce the output
  array(s).
Many areas, such as image processing and scientific computing,
  widely use such a pattern.
%In fact, the convolution operation is a stencil computation.
While the sliding window pattern seems regular and simple,
  it is non-trivial to optimize the stencil computation kernels for performance given its low
  computation-to-communication ratio along and
  complex data dependency patterns.
  %and the often slower memory system that cannot keep up with the computation units.
Even worse,
  a stencil computation kernel can be composed of many stages or iterations
  concatenated with each other,
  which further complicates data dependency and makes communication
  optimizations harder to achieve. 
To overcome these challenges, we developed a stencil compiler named SODA~\cite{iccad18-soda, dac20-soda-cr} with the following customized optimization for the stencil microarchitecture:

\begin{itemize}

\item \textbf{Parallelization support}.
The stencil computation has a large degree of inherent parallelism,
  including both \textit{spatial parallelism}, i.e.,
  parallelism among multiple spatial elements within a stage,
  and \textit{temporal parallelism}, i.e.,
  parallelism among multiple temporal stages. 
SODA exploits both fine-grained spatial parallelism and coarse-grained
  temporal parallelism with perfect data reuse,
  by instantiating multiple processing  elements (PE) in each stage and
  concatenating multiple stages together on-chip, respectively.
Figure~\ref{fig:soda-overview} illustrates the microarchitecture overview of
  a SODA DSA with three fine-grained PEs per stage and
  two coarse-grained stages.
Within each stage, multiple (three in Figure~\ref{fig:soda-overview})
  PEs can read necessary inputs from the reuse buffers and
  produce outputs in parallel, exploiting spatial parallelism.
Meanwhile,
  Stage~1 can directly send its outputs to Stage~2 as inputs,
  further exploiting temporal parallelism.

\item \textbf{Data reuse support}.
The sliding window pattern makes it possible to reuse input data for multiple
  outputs and reduce memory communication.
CPUs (and to a less extent GPUs) are naturally at a disadvantage for such data reuse due
  to their hardware-managed, application-agnostic cache systems.
FPGAs and ASICs, however, can customize their data paths to reduce the memory
  traffic and achieve the optimal data reuse (in terms of the least amount of off-chip data movement and the smallest possible on-chip memory size) without sacrificing the parallelism as seen in ~\cite{iccad18-soda}.
Figure~\ref{fig:soda-overview} shows the reuse buffers in a SODA DSA,
  which read three new inputs in parallel,
  keep those inputs that are still needed by the PEs in the pipeline,
  and discard the oldest three inputs which are no longer required.
Thus, the SODA DSA accesses both the inputs and outputs in a streamed fashion,
  achieving the least amount of off-chip data movement.
Moreover,
  SODA can generate reuse buffers for any number of PEs with the
  %constant
  smallest possible on-chip memory size, independent of the number of PEs used, ensuring high scalability.

\item \textbf{Computation reuse support}.
Computation operations are often redundant and can be reused for
  stencil kernels.
As an example,
  for a $17\times17$ kernel utilized in
  calcium image stabilization~\cite{fpga19-folding},
  we can dramatically reduce the number of multiplication operations from 197
  to only 30, while yielding the same throughput.
However, computation reuse is often under-explored,
  since most stencil compilers are designed for instruction-based processors where the parallelization and communication reuse have more impact on the performance, while the computation reuse is often just a by-product~\cite{sc19-stencil}.
For DSAs,
  we can fully decouple the computation reuse from parallelization and
  data reuse via data path customization.
SODA
%\AS{maybe it's better to cite both papers together in the beginning}
%provides two comprehensive design-space exploration
  algorithms:
  1) optimal reuse by dynamic programming (ORDP) that can fully explore the
  trade-off between computation and storage when the stencil kernel contains
  up to 10 inputs,
  2) heuristic search–based reuse (HSBR) for larger stencils to find
  near-optimal design points within limited time and memory constraints.

\end{itemize}
\vspace{-0.1in}

We developed the SODA compiler to automatically generate a DSA to consider optimizations in these three dimensions.
SODA utilizes a simple domain-specific language (DSL) as input.
As an example, Code~\ref{code:soda} shows a blur filter written in SODA DSL.
We will discuss later in Section~\ref{sec:heterocl} that higher-level
  domain-specific languages can generate the SODA DSL, further reducing the
  threshold for software programmers and domain experts to benefit from DSAs.
SODA can automatically explore a large solution space for a stencil DSA,
  including unroll factor, iterate factor, tile sizes, and computation reuse.

\begin{lstlisting}[
  caption=Blur filter written in SODA~\cite{iccad18-soda}.\label{code:soda},
  escapechar=|,
  otherkeywords={kernel,unroll,iterate,factor,input,local,output},
  morecomment={[l]\#},
  ]
kernel: blur
unroll factor: 16              # Spatial parallelism
iterate factor: 1              # Temporal parallelism
input float: image(3000, *) # Tile size, which decides reuse buffer size
local float: blur_x(0, 0) = (image(0, 0) + image(1, 0) + image(2, 0)) / 3
output float: blur_y(0, 0) = (blur_x(0, 0) + blur_x(0, 1) + blur_x(0, 2)) / 3
\end{lstlisting}

Experimental results show that SODA, as an automated accelerator design
  framework for stencil applications with scalable parallelization,
  full communication-reuse, and effective computation reuse,
  achieves a $10.9\times$ average
  speedup over the CPU and a $1.53\times$ average speedup over the
  GPU~\cite{dac20-soda-cr}.
This is achieved by over $200\times$ lines of HLS code generated from SODA DSL,
  with 34\% of the lines of code being pragmas.
Such extensive amount of code required by the domain-specific architectural
  optimizations was only possible with a fully automated DSA compiler framework.
\rev{
Compared to SODA, a very recent  work named ClockWork~\cite{fccm21-clockwork}
  achieves less resource consumption by compiling the entire application into
  one flat statically scheduled module, but
  at the cost of long compilation time and scalability to the whole chip.  This presents an interesting trade-off.
A common limitation of the approach taken by both SODA and ClockWork
  is that both cannot create a quickly reconfigurable overlay accelerator
  to accommodate different stencil kernels.
Both have to generate different accelerators for different stencil patterns,
  which may take many hours.
It will be interesting to investigate to see if it is possible to come up with
  a stencil-specific programmable architecture to allow runtime reconfiguration
  for acceleration of different stencil applications.
%faster implementation methodologies to mitigate this issue~\cite{fpga22-rapidstream}.
}

\subsection{Automated Program Transformation and Pragma Insertion} \label{sec:autodse}

\label{sec:automated-program-transformation}

For general applications that do not match the predefined computation patterns (such as systolic arrays and stencil computations), we carry out an automated local program transformation and pragma insertion based on the input C/C++ code. 
The recently developed Merlin Compiler~\footnote{Recently open-sourced by Xilinx at \url{https://github.com/Xilinx/merlin-compiler}}~\cite{merlin_islped} addresses this need partially by providing higher-level pragmas.  
The programming model of the Merlin Compiler is similar to that of OpenMP~\cite{dagum1998openmp}, which is commonly used for multi-core CPU programming. Like in OpenMP, it optimizes the design by defining a small set of compiler directives in the form of pragmas. Codes~\ref{code:omp} and~\ref{code:merlin} show the similarity of the programming structure between the Merlin Compiler and OpenMP.

\begin{center}
\begin{minipage}[t]{0.5\linewidth} 
\begin{lstlisting}[caption=OpenMP\label{code:omp}]
#pragma omp parallel for num_threads(16)
for (int i = 0; i < N; i++) {
    c[i] += a[i] * b[i]; }
\end{lstlisting}
\end{minipage}
\quad
\begin{minipage}[t]{0.45\linewidth}
\begin{lstlisting}[caption=Merlin Compiler\label{code:merlin}]
#pragma ACCEL parallel factor = 16
for (int i = 0; i < N; i++) {
    c[i] += a[i] * b[i]; }
\end{lstlisting}
\end{minipage}
\end{center}

% [language=C,caption=OpenMP for Multi-core CPUs\label{code:omp}] 
%\input{codes/merlin}

Table~\ref{tbl:merlin_pragmas} lists the Merlin pragmas for the design architecture transformations. Using these pragmas, the Merlin Compiler will apply source-to-source code transformations and generate the equivalent HLS code with proper HLS pragmas inserted. The \texttt{fg PIPELINE} refers to the case where \textit{fine-grained} pipelining is applied by pipelining the loop and unrolling all its inner loops completely. In contrast, the \texttt{cg PIPELINE} pragma applies \textit{coarse-grained} pipelining by creating double buffers automatically between the pipelined tasks.
\begin{table}[!htb]
% \footnotesize
\centering
\caption{Merlin pragmas with architecture structures.}
\label{tbl:merlin_pragmas}
\begin{tabular}{ccc}
\hline
\multicolumn{1}{|c|}{Keyword}                   & \multicolumn{1}{c|}{Available Options}             & \multicolumn{1}{c|}{Architecture Structure} \\ \hline\hline
\multicolumn{1}{|c|}{PARALLEL}                  & \multicolumn{1}{c|}{factor=\textless{}int\textgreater{}}                    & \multicolumn{1}{c|}{CG \& FG parallelism}   \\ \hline
\multicolumn{1}{|c|}{PIPELINE} & \multicolumn{1}{c|}{mode=cg/fg}                       & \multicolumn{1}{c|}{CG or FG pipeline}          \\ \hline
\multicolumn{1}{|c|}{TILING}                  & \multicolumn{1}{c|}{factor=\textless{}int\textgreater{}}                    & \multicolumn{1}{c|}{Loop Tiling}   \\ \hline
\multicolumn{3}{c}{CG: Coarse-grained; FG: Fine-grained}
\end{tabular}
\end{table}

By introducing a reduced set of high-level pragmas and generating the corresponding HLS code automatically, the Merlin Compiler can make FPGA programming  significantly easier. For example, we can optimize the Advanced Encryption Standard (AES) kernel from the MachSuite benchmark~\cite{reagen2014machsuite} by adding only 2 Merlin pragmas and achieve a $470\times$ speedup compared to when the kernel (without any changes) is synthesized using the Vitis HLS tool. However, the manually optimized HLS code utilizes 37 pragmas and 106 more lines of code to get to the same performance.

Although the Merlin Compiler greatly reduces the solution space when optimizing a kernel, it still requires the programmer to manually insert the pragmas at the right place with the right option, which can still be challenging. To further reduce the DSA design effort, we have developed a push-button design space exploration (DSE), called \textit{AutoDSE}~\cite{sohrabizadeh2020autodse}, on top of the Merlin Compiler. AutoDSE is designed to automatically insert Merlin pragmas to optimize the design based on either performance, area, or a trade-off of the two.

\begin{figure}[!htb]
	\centering
	\includegraphics[width=\linewidth]{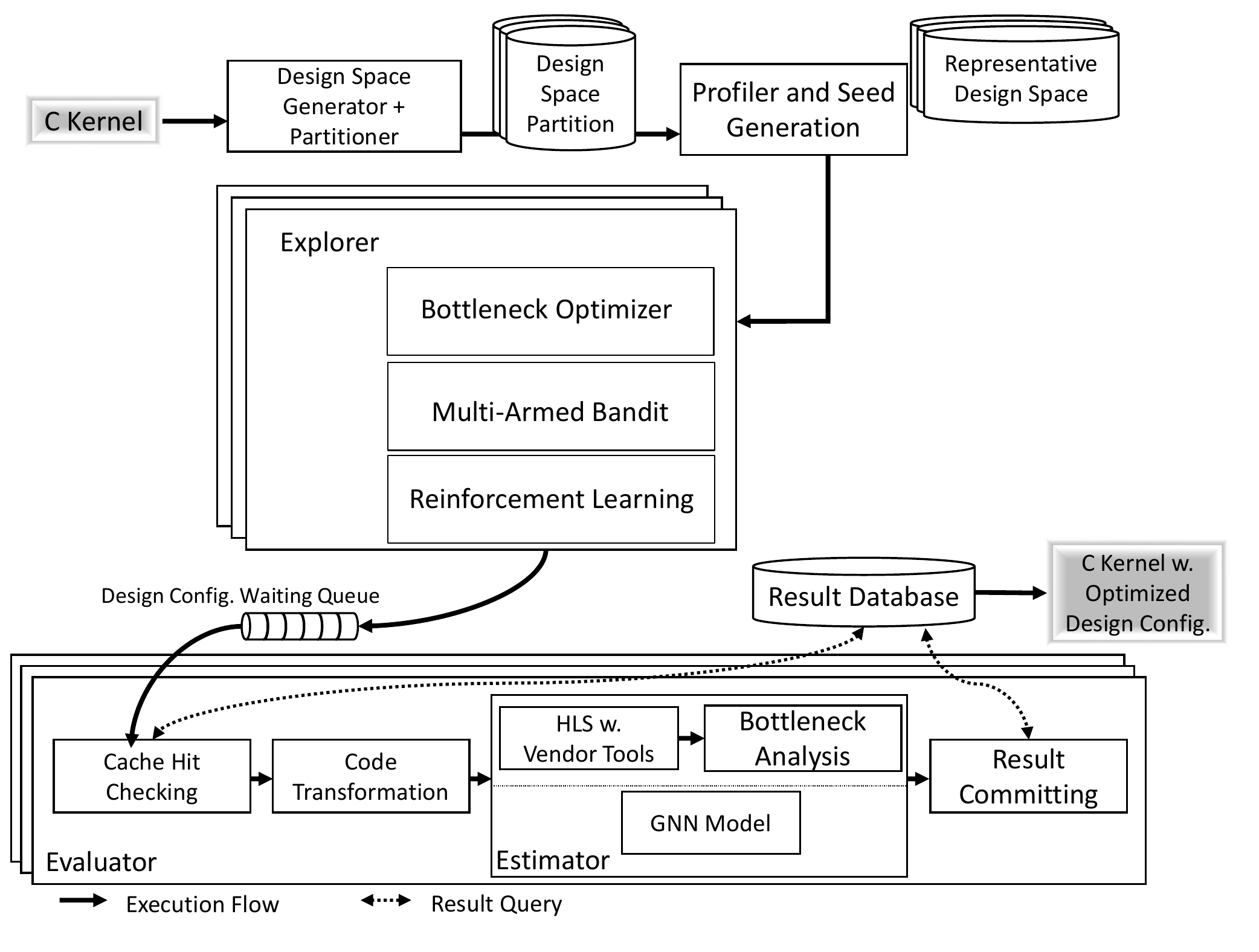} 
	\caption{The AutoDSE framework as in~\cite{sohrabizadeh2020autodse}.}
	\label{fig:autodse}
\end{figure}

As depicted in Figure~\ref{fig:autodse}, AutoDSE takes the C kernel as the input and identifies the design space by analyzing the kernel abstract syntax tree (AST) to extract the required information such as loop trip-counts and available bit-widths. It then encodes the valid space in a grid structure that marks  the invalid pragma combinations. As the design parameters have a non-monotonic impact on both the performance and area, AutoDSE partitions the design space to reduce the chance of it getting trapped in a locally optimal design point. Each partition explores the design space from a different starting design point so that AutoDSE can search different parts of the solution space. %Each partition runs an independent explorer to search for the Pareto-optimal design points, which can be easily parallelized, and all the results are stored in a common database. 
Once the DSE is finished, AutoDSE will output the design point with the highest quality of results (QoR).

AutoDSE is built on a bottleneck-guided optimization that mimics the manual optimization approach to perform iterative improvements. 
%It takes the HLS tool as a black box and, 
At each iteration, AutoDSE runs the Merlin Compiler to get the detailed performance breakdown of the current design solution (estimated after HLS synthesis). Then, it identifies a set of performance bottlenecks, sorted by decreasing latency, and marks each entry as computation or communication-bounded. 
%It then uses the sorted hierarchy paths to output an ordered list of the design parameters based on the bottleneck type. 
Guided by this list of bottlenecks, at each iteration, AutoDSE applies the appropriate Merlin pragma to the code section with the most-promising impact for performance improvement. The experimental results show that using the bottleneck-optimizer, AutoDSE is able to achieve high-performance design points in a few iterations. Compared to a set of 33 HLS kernels in Xilinx Vitis libraries~\cite{vitis-library}, which are optimized manually, AutoDSE can achieve the same performance while utilizing $26.38\times$ fewer optimization pragmas for the same input C programs, resulting in less than one optimization pragma per kernel\footnote{the rest of the pragmas consist of \texttt{STREAM} and \texttt{DATAFLOW} which are not included in the Merlin's original pragmas. We will add them in the future. \rev{Note that the Merlin Compiler can directly work with Xilinx HLS pragmas as well.}}. Therefore, in combination with the Merlin Compiler, AutoDSE greatly simplifies the DSA design effort on FPGAs, making it much more accessible by software programmers who are familiar with CPU performance optimization.
 
{\revision Schafer et al.~\cite{dse-survey} provided a good survey of the prior DSE works up to 2019. They either invoke the HLS tool for evaluating a design point, as in AutoDSE, or develop a model to mimic the HLS tool. Employing a model can speed up the DSE process since we can assess each point in milliseconds instead of several minutes to even hours. However, as pointed out in~\cite{dse-survey}, directly using the HLS tool results in a higher accuracy. AutoDSE has shown to outperform the previous state-of-the-art works. Nevertheless, relying on the HLS tool does slow down the search process considerably and limit the scope of exploration.  
To speedup the design space exploration process, we are developing a \textit{single} graph neural network (GNN)-based model for performance estimation to act as a surrogate of the HLS tool across different applications. Initial experimental results show that a GNN-based model can estimate the performance and resource usage of each design point with high accuracy in milliseconds~\cite{gnn-dse, bai2022improving}}. We are excited by the prospect of applying machine learning techniques to DSA synthesis.

\subsection{Further Raising the Level of Design Abstraction} \label{sec:heterocl}
%{\highlight{(Yuze, 1p)}}- Example: HeteroHalide (via HeteroCL)

%While the advance of HLS (Section~\ref{sec:hls-limitation}),
While architecture-guided optimizations
  (Section~\ref{sec:architecture-guided-optimation})
  and automated program transformation
  (Section~\ref{sec:automated-program-transformation}) make it a lot easier to achieve high-performance DSA designs from C/C++ programs, the software community has introduced various domain-specific languages (DSLs) for better design productivity in certain application domains.
One good example is Halide~\cite{halide}, a widely-used image processing DSL,
  which has the advantageous property of decoupling the algorithm specification from performance optimization (via scheduling statements).
This is very useful for image processing applications,
  because it is difficult and time-consuming for a designer to write image
  processing algorithms while parallelizing execution and optimizing for data
  locality and performance at the same time,
  due to the large number of processing stages and the complex data dependency.
%\rev{Merging the two paragraphs as suggested by Reviewer 2}
However, the plain version of Halide only supports CPUs and GPUs.
% \rev{
%   Removing a sentence that makes this part somewhat awkward.
%   The point has been covered briefly in the first sentence of this paragraph.
% }
% while the Halide community has been growing rapidly in recent years (received  over 4,500 stars on GitHub),
There is no way to easily synthesize the vast number of Halide programs to DSAs on FPGAs.
The direct and traditional way is to rewrite programs in hand-optimized RTL code or HLS C code, which is very time-consuming.
%Although C/C++-based high-level synthesis (HLS) raises the design abstraction level to untimed behavior description with automated scheduling, pipelines, and resource sharing~\cite{cong11}, it still requires circuit-design expertise to get efficient designs,  which becomes a high threshold for software programmers.
%We would like to lower the bar as much as possible,
Our goal is to develop an efficient Halide-based compiler to DSA implementations on FPGAs.
%yet still benefit from the latest advances of domain-specific architectural
%  optimizations (e.g., SODA~\cite{iccad18-soda}) and generate DSAs with
%  state-of-the-art quality of result.
%We could generate SODA DSL directly from Halide, but doing so will overly
%  restrict the input to stencil patterns only,
%  and we would like to support non-stencil kernels as well.

Our approach is to leverage the recently developed
  HeteroCL~\cite{fpga19-heterocl} language as an intermediate
  representation (IR).
As a heterogeneous programming infrastructure, HeteroCL provides a Python-based
  DSL with a clean programming abstraction that
  decouples algorithm specification from three important types of hardware
  customization in \textit{compute}, \textit{data types},
  and \textit{memory architectures}.
HeteroCL further captures the interdependence among these different
  customizations, allowing programmers to explore various
  performance/area/accuracy trade-offs in a systematic and productive way.
In addition, HeteroCL produces highly efficient  hardware implementations for a
  variety of popular workloads by targeting spatial architecture templates
  (Section~\ref{sec:architecture-guided-optimation}) including systolic arrays
  (Section~\ref{sec:autosa}) and stencil (Section~\ref{sec:soda}).
HeteroCL allows programmers to explore the design space efficiently
  in both performance and accuracy by combining different types of
  hardware customization and targeting spatial architectures,
  while keeping the algorithm code intact.

\begin{figure}[!ht]
  \centering
  \includegraphics[width=\linewidth]{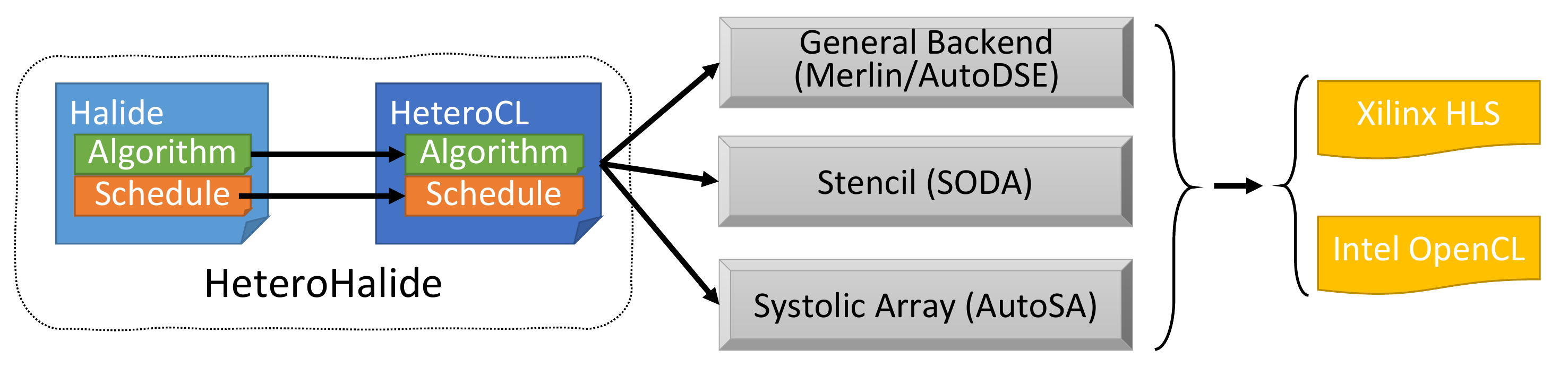}
  \caption{HeteroHalide~\cite{fpga20-heterohalide} overall workflow.}
  \label{fig:heterohalide-overview}
\end{figure}

On top of HeteroCL, we developed HeteroHalide~\cite{fpga20-heterohalide},
  an end-to-end system for compiling Halide programs to DSAs.
\rev{As a superset of Halide,}
HeteroHalide leverages HeteroCL~\cite{fpga19-heterocl} as an intermediate
  representation (IR) to take advantage of its vendor neutrality,
  great hardware customization capability,
  and the separation of algorithms and scheduling.
The multiple heterogeneous backends (spatial architectures) supported by
  HeteroCL makes HeteroHalide able to generate efficient hardware code according
  to the type of applications.
Figure~\ref{fig:heterohalide-overview} shows the overall workflow of
  HeteroHalide.

HeteroHalide greatly simplifies the migration effort from plain Halide,
  since the only requirement is moderate modifications on the scheduling part,
  not algorithm.
HeteroHalide automatically generates HeteroCL~\cite{fpga19-heterocl} code,
  making use of both algorithm and scheduling information specified in a Halide
program.
Code~\ref{code:heterohalide} demonstrates a blur filter written in HeteroHalide.
By changing only the scheduling part of the Halide code,
  HeteroHalide can generate highly efficient FPGA accelerators that outperform
  28 CPU cores by $4.15\times$ on average on 10 real-world Halide applications,
  including 8 applications using the stencil backend,
  1 using the systolic-array backend,
  and 1 using the general backend~\cite{fpga20-heterohalide}.
HeteroHalide has achieved this not only by taking advantage of HeteroCL,
  but also by adding hardware-specific scheduling primitives to plain Halide.
For example, when we compile plain Halide,
  the scheduling is applied directly at IR level using
  \textit{immediate transformation} (Line~\ref{line:immediate}).
This may result in loss of information during such transformations,
  which could prevent lower-level compilers from applying optimizations.
We create extensions on Halide schedules,
  allowing some schedules to be lowered with annotations, using
  \textit{lazy transformation} (Line~\ref{line:lazy}).
By adding this extension to Halide,
  HeteroHalide can generate specific scheduling primitives at the HeteroCL
  backend level,
  thus emitting more efficient accelerators,
  and an image-processing domain expert will be able to leverage DSAs by just
  changing the scheduling part of their existing Halide code.
For the following sample Halide code, HeteroHalide is able to generate
  1455 lines of optimized HLS C code with 439 lines of pragmas,
  achieving $3.89\times$ speedup over 28 CPU cores
  \rev{using only one memory channel of the AWS F1 FPGA}~\cite{fpga20-heterohalide}.
% ~jiajieli/app_halide2heterocl/blur/codegen/blur_kernel.cpp
\rev{
Using all four memory channels,
  HeteroHalide can outperform the Nvidia A100 GPU
  (that is $2.5\times$ more expensive on AWS) by $1.1\times$
  using schedules generated by
  the Li2018 Halide auto-scheduler~\cite{tog18-li2018}. We expect to see more gains when sparse computation is involved.
}

In general, HeteroHalide demonstrated a promising flow of compiling high-level DSLs via HeteroCL to FPGAs for efficient DSA implementation. In addition, the newly emerged MLIR compilation framework~\cite{mlir} is also promising as an alternative intermediate representation, which we plan to explore in the future.
More opportunities to improve HLS are discussed in
a recent keynote invited paper in~\cite{cong2022fpga}.

\vspace{0.1in}
\begin{lstlisting}[
  caption=Blur filter written in HeteroHalide~\cite{halide}.\label{code:heterohalide},
  escapechar=|,
  ]
  // Algorithm, same as plain Halide
  Func blur_x("blur_x"), blur_y("blur_y");
  blur_x(x, y) = (image(x, y) + image(x+1, y) + image(x+2, y)) / 3;
  blur_y(x, y) = (blur_x(x, y) + blur_x(x, y+1) + blur_x(x, y+2)) / 3;

  if (for_hardware) {
    blur_x.lazy_unroll(x, 16); // Schedule for hardware, added for HeteroHalide|\label{line:lazy}|
  } else {
    blur_x.unroll(x, 16);         // Schedule for software, same as plain Halide|\label{line:immediate}|
  }
\end{lstlisting}

\section{Concluding Remarks} \label{sec:conclusion}
In this article, we show that with architecture-guided optimizations, automated design space exploration for code transformation and optimization, and support of high-level DSLs, we can provide a programming environment and compilation flow  that is friendly to software programmers and empower them to create their own DSAs on FPGAs with efficiency and affordability. This is a critical step towards democratization of customized computing.

The techniques presented in this article are not limited to existing commercially available FPGAs, which were heavily influenced  by communication, signal processing, and other industrial applications that dominated the FPGA user base in the early days.  To address the growing needs for computing acceleration, a number of features have been added, such as the dedicated floating processing units in the Intel's Arria-10 FPGA family and the latest AI processor engines in the Xilinx's Versal FPGA family. We expect this trend will continue, for example, to possibly incorporate the concept of coarse-grained reconfigurable arrays (CGRAs)~\cite{cgra} to reduce the overhead of fine-grained programmability, and greatly reduce the long physical synthesis time suffered by existing FPGAs.  Our compilation and optimization techniques are readily extensible to such coarse-grained programmable fabrics.  For example, we have an ongoing project of applying our systolic array compile to the array of AI engines of the Versal FPGA architecture and adding CGRA overlays to exisitng FPGAs~\cite{overgen}. 

In their 2018 Turing Award lecture entitled ``A golden age for computer architecture'', Hennessy and Patterson concluded that ``the next decade will see a Cambrian explosion of novel computer architectures, meaning exciting times for computer architects in academia and in industry''~\cite{hennessy2019new}, with which we fully agree.  Our research aims at broadening the participation of this exciting journey, so that not only computer architects, but also a large number of performance-oriented software programmers can create their own customized architectures and accelerators on programmable fabrics to achieve a significant performance and energy efficiency improvement. We hope that this article can stimulate more research in this direction.

%\vspace{-0.1in}
\begin{acks}
We would like to thank Marci Baun for editing the paper. 
This work is supported in
part by CRISP, one of six centers in JUMP, a Semiconductor
Research Corporation (SRC) program co-sponsored by DARPA, the CAPA award jointly funded by NSF (CCF-1723773) and Intel (36888881), and CDSC industrial partners (\url{https://cdsc.ucla.edu/partners/}).
\end{acks}

\bibliographystyle{ieeetr}
\bibliography{main}

\end{document}